# Shapes of Direct Cortical Responses vs. Short-Range Axono-Cortical Evoked Potentials: The Effects of Direct Electrical Stimulation Applied to the Human Brain

Clotilde Turpin[1], Olivier Rossel[1], Félix Schlosser-Perrin[1], Sam Ng[4], Riki Matsumoto[2], Emmanuel Mandonnet[3], Hugues Duffau[4,+], François Bonnetblanc[1,+,*]

[1]CAMIN Team, INRIA, Université de Montpellier, France

[2]Division of Neurology, Kobe University Graduate School of Medicine, Japan

[3]Département de Neurochirurgie, Centre Hospitalier Universitaire, Hôpital Lariboisière, Paris, France

[4]Département de Neurochirurgie, Centre Hospitalier Universitaire de Montpellier Gui de Chauliac, Montpellier, France

*Corresponding author. E-mail address: francois.bonnetblanc@inria.fr (F. Bonnetblanc).

+Both authors supervised this work.

**Abstract:**

*Objective:* Direct cortical responses (DCR) and axono-cortical evoked potentials (ACEP) are generated by electrically stimulating the cortex either directly or indirectly through white matter pathways, potentially leading to different electrogenic processes. For ACEP, the slow conduction velocity of axons (median $\approx$ 4 m.s$^{-1}$) is anticipated to induce a delay. For DCR, direct electrical stimulation (DES) of the cortex is expected to elicit additional cortical activity involving smaller and slower non-myelinated axons. We tried to validate these hypotheses.

*Methods*: DES was administered either directly on the cortex or to white matter fascicles within the resection cavity, while recording DCR or ACEP at the cortical level in nine patients.

*Results*: Short but significant delays ($\approx$ 2 ms) were measurable for ACEP immediately following the initial component ($\approx$ 7 ms). Subsequent activities ($\approx$ 40 ms) exhibited notable differences between DCR and ACEP, suggesting the presence of additional cortical activities for DCR.

*Conclusion:* Distinctions between ACEPs and DCRs can be made based on a delay at the onset of early components and the dissimilarity in the shape of the later components (>40 ms after the DES artifact).

*Significance:* The comparison of different types of evoked potentials allows to better understand the effects of DES.

## 1. Introduction

The measurement of potentials evoked (EP) by direct electrical stimulation (DES) during brain surgery is emerging as a method to identify structural or anatomical connectivity in real-time, while aiming to preserve it (Boyer et al., 2021a; Mandonnet et al., 2016; Matsumoto et al., 2007, 2004; Rossel et al., 2023; Schlosser-Perrin et al., 2023; Vincent et al., 2020; Yamao et al., 2014a). This innovative approach facilitates the recording and observation of various responses in the operating room, including (i) the direct cortical response (DCR: DES and recording at the same cortical site, on the same gyrus), (ii) cortico-cortical evoked potentials (CCEP: DES and recording on distant cortical sites), and (iii) axono-cortical evoked potentials (ACEP: DES on white matter tracts with corresponding cortical recording). It is now becoming more and more routine to record these three types of responses in the same patients (Boyer et al., 2021a), providing a more comprehensive understanding of the nature of the collected EPs and the general effects of DES.

While DCR and ACEP are generated through different means by electrically stimulating the cortex either directly or indirectly through white matter tracts, their waveforms exhibit minimal variation, and these two types of EP share identical initial components (Boyer et al., 2021b; Rossel et al., 2023). Specifically, these responses consist of an early positive component, denoted as P0, occurring after the DES artifact and before 7-10 ms when measurable. Subsequently, a robust negative deflection, labeled N1, is observed with its peak occurring between 10-25 ms. As DES initially activates larger neuronal elements (Blair and Erlanger, 1933; McNeal, 1976; Rattay, 1999; Ruch TC and Patton HD, 1982), and considering its early occurrence, P0 is hypothesized to reflect a summation of highly synchronized action potentials (AP). This interpretation is not novel; (Sugaya et al., 1964) already proposed that "the several brief positive spikes which initiate DCR when the evoking stimulus is strong were attributed to the all-or-none discharge of neuron somata situated in the cortical depth." On the contrary, the latency of N1 suggests a direct association with the summation of postsynaptic potentials, particularly excitatory ones (EPSP, characterized by slower dynamics) (Adrian, 1936; Chang, 1951; Li and Chou, 1962).

Nowak and Bullier, 1998a, b even suggest that, for the largest neuronal elements activated initially, axons, rather than somas, generate APs when electrical stimulation is applied to the cortex. As illustrated in Figure 1, we hypothesized that there are substantial similarities in the electrogenesis of DCRs and ACEPs: both responses entail initially the activation of the larger afferent and efferent axons within the cortical column, leading to (i) the occurrence of the P0 component and (ii) the subsequent integration of these volleys of APs at the dendritic level through EPSP, thereby giving rise to the N1 component. A notable difference between DCR and ACEP is associated with the delay between axonal stimulation in the case of ACEP and its propagation to the cortical layer (See (1) in Figure 1). It is well-established that conduction speed depends on the size of myelinated axons, as per repeated empirical observations: velocity (m.s$^{-1}$) = 6 x Diameter (µm) (Gasser and Grundfest, 1939; Hursh, 1939; Liewald et al., 2014; Rushton, 1951; Waxman and Bennett, 1972; Waxman and Swadlow, 1977). Given that the diameter of myelinated axons in the white matter ranges from [0.1;9] µm with a median around 0.7 µm (Liewald et al., 2014), this approach suggests that a DES applied 1 cm further on the white pathways could induce delays inferior of around 2.4 ms for ACEP in comparison to DCR, due to activation of bigger axons with DES.

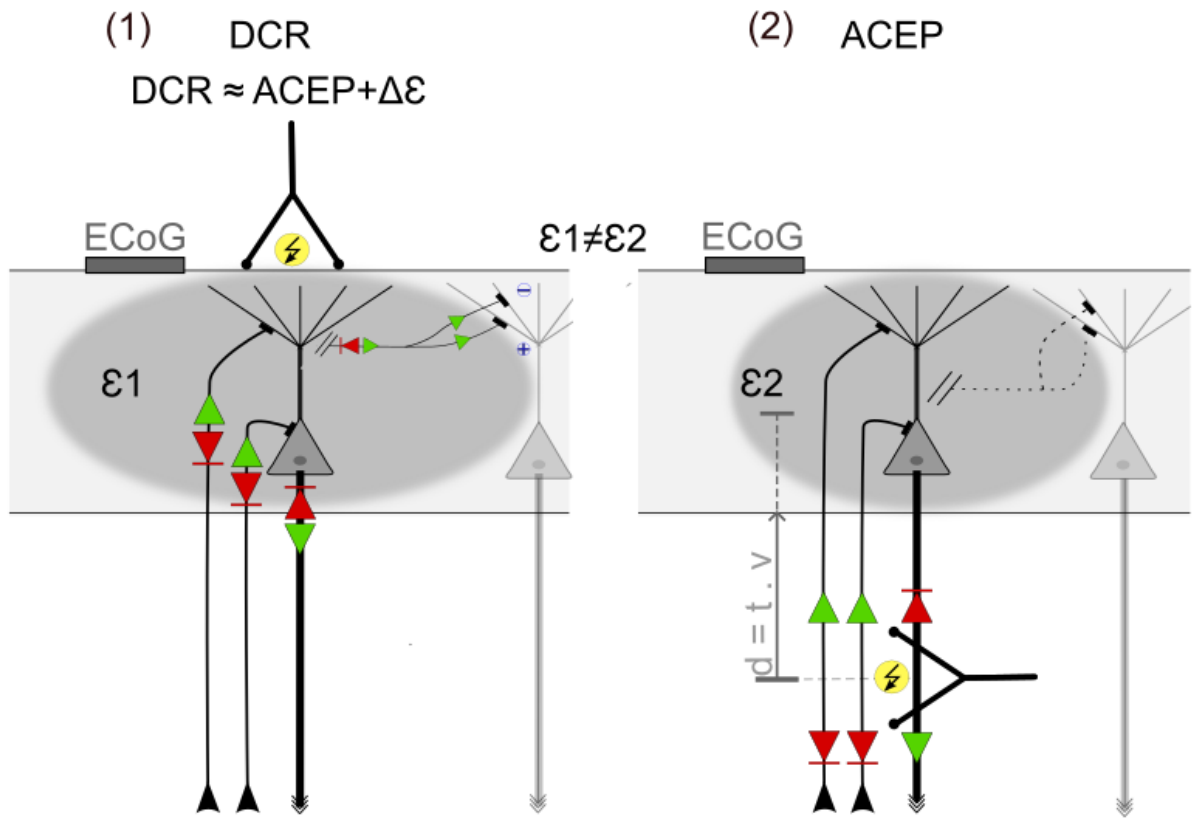

**Figure 1.** Scheme of the hypothetical electrogenesis provoked by the application of DES to different sites and generating the two evoked responses. (1) For DCR, DES was applied at the cortical surface and in the same gyrus. (2) For ACEP, DES was applied on the white matter pathways. Due to DES, axonal activities were generated ortho- (green arrows) and antidromically (red arrows) in radial axons (both for DCR and ACEP) and tangential smaller intra-cortical axons (DCR only). $\varepsilon_1$ and $\varepsilon_2$ are defined as the secondary cortical activities. Main processes involved in the electrogenesis of the two types of EP are identical. However, two hypotheses possibly distinguishing the electrogenesis process involved in DCR vs. ACEP are presented here: (1) additional intra-cortical activity involved in DCR and (2) phase delay for ACEP with a more distant site of DES. These slight differences could be captured by ECoG recordings.

On the contrary, in the context of DCR, as proposed in (1) of Figure 1, DES applied directly to the cortex may also induce activities in intra-cortical axons and dendrites, contrasting with the "more natural" activations observed through white matter pathways, as is the case for ACEP. It can be hypothesized that the axons, being considerably smaller in size at the cortical level and lacking myelination (Markram et al., 2015), will exhibit slower dynamics in these additional activations, consequently becoming observable at later stages in the evoked responses.

We posit that a measurable delay will exist at the level of the early components of ACEP compared to those of DCR. Additionally, distinct activity is expected to be observable later due to direct activation at the cortex level in the case of DCR (refer to (1) and (2) in Figure 1).

## 2. Methods

Nine patients underwent awake brain surgery for tumor resection. They were operated on in Montpellier University Medical Center. The study (UF 965, n° 2014-A00056-43) was approved by the local ethics committee. All patients gave informed and written consents to participate in the study.

### *2.1. Anaesthetic protocols*

Patients underwent "Intermittent general anesthesia with controlled ventilation for asleep awake asleep brain surgery" (Deras et al., 2012). Briefly, general anesthesia was induced by intravenous infusion of DIPRIVAN (propofol) and remifentanyl, which are short-acting medications and opioid analgesic drugs, respectively, that have a rapid onset and recovery time. During the first phase of the surgery, patients were systematically under general anesthesia and a laryngeal mask airway was used. The target concentrations were 6 mg ml$^{-1}$ and 6 ng ml$^{-1}$ for propofol and remifentanyl. During the second phase of the surgery, all anesthetic drugs were stopped, and the laryngeal mask was removed. Finally, following tumor resection, patients underwent a second general anesthesia for dural, skull, and skin closures. At this stage, a laryngeal mask of tracheal intubation was used, depending on the anesthesiologist's decision. When tracheal intubation was preferred, the targets were 12 mg ml$^{-1}$ and 12 ng ml$^{-1}$, for propofol and remifentanil, respectively.

### *2.2. Intraoperative functional mapping*

Standard cortical and white matter mapping were performed on all awake patients. Intraoperative functional mapping was carried out using 60 Hz direct electrical stimulation (DES) while patients performed cognitive and motor tasks. Functional sites identified during the operation were tagged with numbered labels placed on the brain surface. A manual bipolar probe was used to deliver an adjustable biphasic current square wave pulse with the Nimbus i care (Innopsys, Toulouse, France) at frequencies of 60 Hz. The intensity was set between 1.5 and 3 mA, with a single pulse duration (PW) of 1 ms. The intensity was increased in steps of 0.5 mA until the stimulation elicited a clinical response, but without inducing seizures. Before the resection, DES was performed on the entire exposed cortical areas, with the probe being displaced every 5 mm in two orthogonal directions. Then, DES was also applied at during the resection on the white matter pathways to identify critical functional sites. The duration of the effective stimulation was determined based on the mapped function, with 1 s used to induce positive motor and sensory responses and up to 4 s used to inhibit cognitive ones. Several stimulation conditions were tested and repeated twice non consecutively at each stimulation site. If a functional disturbance was consistently induced, the surgeon avoided resecting the stimulated area.

### *2.3. Post-resection Electrocorticographic (ECoG) recordings*

Electrocorticographic (ECoG) recordings were performed under general anesthesia at the end of the resection. Cortical and subcortical sites (white matter pathways) were stimulated again with low frequency Direct Electrical Stimulation (DES) at 9 Hz (except for Patient 2 and 8, $f_{DES}$=3 Hz), with pulse duration (PW) of 1 ms, in order to record DCR and ACEP respectively. ECoG data were recorded after tumor resection using platinum contact strips with a radius of 2.5 mm and 10 mm spacing (DIXI, France) placed on the surface of the brain. ECoG signals were recorded using a referenced (common or monopolar) mode, with a sampling rate of 19.2 kHz (g. HIAMP, G.tec, Austria). Ground and reference electrodes (Medtronic DME 1004) were placed on the acromion and ipsilesional mastoid, respectively. Multiple channels were measured for each patient, and signals were recorded without filtering. Post-processing (using Matlab® software) involved removing artifacts (replaced by interpolation, during stimulation duration plus the following 4 ms) then 50 Hz noise cancelling and filtering the data between 1 Hz and 1000 Hz. The procedure for identifying and removing the 50 Hz noise involved selecting a channel with higher noise, building an averaged pattern representative of the noise, selecting another channel where the averaged pattern repeated, extracting a second averaged pattern specific to this channel, and subtracting it. This process was repeated for all channels of interest, taking advantage of the synchronous nature of 50 Hz noise across all channels. For each channel, mean traces of EPs were obtained by (epoching and base line correction) averaging ECoG signals synchronized to the DES. Each stimulation train contained between 7 and 132 pulses, depending on the stimulation rate (i.e. f = 3 or 9 Hz) and duration (varying between 0.8s and 21.4s; Patient1: f=9Hz, T= [9.3:13.8]s, n=[84:124] stimulations; Patient2: 3Hz, [3.7:14]s, [11:42]; Patient 3: 9Hz, [2.6:4.9]s, [23:44]; Patient 4: 9Hz, [3.2:8.6]s, [29:77]; Patient 5: 9Hz, [0.8:21.4]s, [7: 193]; Patient6: 9Hz, [2.4:7.7]s, [22 : 69]; Patient7: 9Hz, [3.6:5.1]s, [32:46]; Patient 8 : 3Hz, [7.3: 5.3]s, [22:46]; Patient 9: 9Hz, [6.2:19.6]s, [56:176]).

The baseline of each stimulus was defined using the mean value of the 5 last milliseconds preceding each stimulus artifact. The stability and repeatability of EPs over time were checked to control for possible adaptation induced by the stimulation frequency. The locations of DES sites, DCR and ACEP evocations are represented on Figure 2 (see Figure 2). Importantly, the precise method for signal processing is detailed in the supplementary material/appendix attached to the present article.

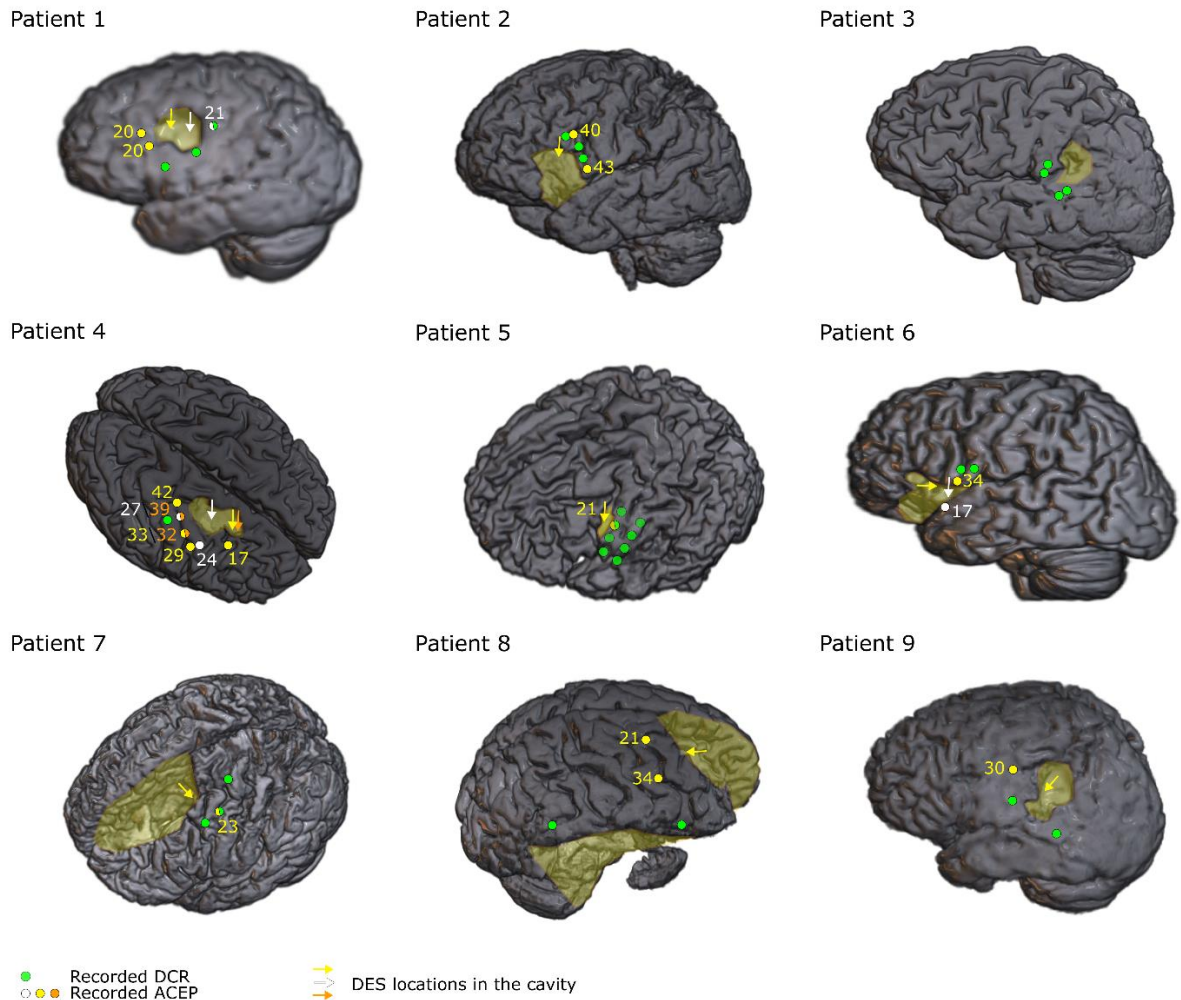

**Figure 2.** Localization of the craniotomy showing the cavity, the ECoG of interest (circles) and DES positioning (arrows) in the 3D post-operative brain for each patient. ECoG were recording DCR and ACEP when colored in green and yellow/red respectively or both. DES was applied on the sites identified by the yellow/red arrows for ACEP only. For DCR DES was applied very close to each ECoG colored in green. For ACEP, estimated distances (in mm) between the DES sites and the recording ECoG were reported in yellow or red.

Overall, 84 DES were considered for all patients. DCR (n= 55) and ACEP (n= 29) with an amplitude of less than 100 μV were not taken into account for comparison.

### 2.4. Metrics of the waveforms

Most signals have a classical waveform with a positive early component (< 10 ms), called the 'P0', followed by a second negative component (< 30 ms), called the 'N1'. These two components can be followed by another positive part occurring between 40 and 100 ms which can be described as the 'after-positivity'. For ACEP, these polarities can be reversed if the ECoG is located on the other side of the sulcus with respect to the generator/ dipole (Rossel et al. 2023). Various parameters for shape, temporality and amplitude were measured on mean EPs (by averaging various epoch of the same train of DES) for comparisons purposes. For sake of clarity, only those showing significant differences were presented here (see Figure 3A). In particular, we investigated whether the relaxation of the N1 was monotonic or if it was characterized by an after-positivity after low-pass filtering (cutoff frequency = 200 Hz). We thus computed the minimum value of the time derivative of DCR and ACEP traces after the N1 peak ($Min(dDCR/dt)_{[50:80]ms}$ and $Min(dACEP/dt)_{[50:80]ms}$): a positive value indicated that the N1 component came back to zero with monotonic positive slopes, whereas a negative value indicated that an after-positivity followed the N1 component.

### 2.5. Frequency content analysis

A comparative analysis of frequency between ACEP and DCR was also performed to discern variations in the presence of gamma activity (GA) during the N1 period. For each train of stimulation, a Short-Time Fourier Transform (STFT) was applied using a 20 ms Hann window centered every 5 ms. To determine the variations of the frequency power in decibels (dB) for each residue, the log transformed power change in each frequency was compared with the power at baseline (namely from -35 to -5ms before the DES artefact using the results of the time-windows centered from -25 to -15ms). The power changes of one frequency, 50Hz, representative of the GA, were computed from the residues of the same train of DES, for various time windows. This mean GA power value was extracted for several time points and averaged for each patient for comparisons between the ACEP and DCR responses.

### 2.6. Statistical analyses

For each patient, various metrics and parameters were averaged together and compared with a paired t-test (after non-normal distribution was rejected with a shapiro-wilk test).

## 3. Results.
### 3.1. Waveforms comparison

Importantly in this study, responses were positive up and negative down.

Figure 3B shows average traces of DCR and ACEP. It reveals that (i) the beginning of the N1 responses (t_zc1; see inset) looked delayed in the ACEP, the width of the DCR ($W_{N1}$ and $WHQ_{N1}$) was greater for the DCR and (ii) the after-positivity was more pronounced for ACEP ($Area_{[40:100]ms}$; $Min(dDCR/dt)_{[50:80]ms}$ ; $Min(dACEP/dt)_{[50:80]ms}$). These three observations were confirmed statistically as reported in Table 1.

| Metrics | DCR Mean value ± std | ACEP Mean value ± std | p-value | Test score | Test |
|---|---|---|---|---|---|
| t_zc1 (ms) | 5.49 ± 1.20 | 7.29 ± 2.24 | 0.0556 | -2.29 | Student test |
| t_zc2 (ms) | 62.73 ± 14.18 | 41.86 ± 7.21 | 0.0064 | 3.84 | Student test |
| $W_{N1}$ (ms) | 57.24 ± 13.68 | 34.58 ± 6.64 | 0.0025 | 4.59 | Student test |
| $WHQ_{N1}$ (ms) | 32.86 ± 7.69 | 24.71 ± 1.72 | 0.0373 | 2.56 | Student test |
| $Area_{[40:100]ms}$ (μV.ms) | -1097.86 ± 1831.87 | 1487.22 ± 1454.52 | 0.0201 | -3.00 | Student test |
| $Min(dDCR/dt)_{[50:80]ms} > 0$ | -0.61 ± 1.42 | | 0.8819 | -1.28 | Student test |
| $Min(dACEP/dt)_{[50:80]ms} < 0$ | | -1.30 ± 0.89 | 0.0022 | -4.13 | Student test |

**Table 1.** Mean values and statistical results for the various metrics of the evoked responses.

$t_{zc\_1}$ : time to the zero-crossing between P0 and N1. $t_{zc\_2}$ : time to the zero-crossing at the end of the N1. $W_{N1}$: width of the N1. $WHQ_{N1}$: width of the N1 at a quarter of the height. $Area_{[40:100]ms}$: Signed area under the curve between 40 and 100 ms. $Min(dDCR/dt)_{[50:80]ms}$ and $Min(dACEP/dt)_{[50:80]ms}$ minimum of the curve slope between 50 and 80 ms. Comparisons were made with respect to zero for these two variables, in order to assess whether the relaxation slope of the N1 was monotonic. Please see also Figure 3A for a graphical illustration of the different variables.

### 3.2. Later N1 beginning for ACEP and estimate of conduction velocity

Interestingly, as hypothesized and as shown in Table 1 and Figure 3C, the initiations of the N1 components (t_zc1, see inset) occurred later (one-tailed t-test) for ACEP in comparison to DCR with a significant delay in around 1.8 ms ± 2.2 on average (median: 2.4 ms) and values between [-1.71: 5.15] ms.

For ACEP, 3D direct Euclidian distances be tween the DES sites and the localization of the ECoG recordings were computed and reported on Figure 2. It gives a mean conduction velocity of mean: 5.9± 43.6, (median 8.38 m.s$^{-1}$) when estimate for each patient were averaged together.

In patients 1, 5 and 7 there are identical sites shared for ACEP and DCR. We also compared the delay of ACEP to DCR shown for those very traces (see figure 3G). For each patient, we associated the two traces from the recording sites common to both the DCR and ACEP. We performed a rank test (Wilcoxon-Mann-Whitney test) to compare the zero-crossing delay measured for each train of DES showing DCR versus ACEP recorded at the same sites. We also observed a delay in zero-crossing for ACEPs, consistent with the within group results. For each case, the measurements in the trains were performed after applying a 3rd-order low-pass filter (without phase lag: function filtfilt in Matlab®) with a cutoff frequency of 250 Hz, between 2 ms after the artifact and the end of the signal.

### 3.3. N1 duration is superior for DCR

The results show that the duration of the N1 ($W_{N1}$) is greater for the DCR. Interestingly, however, this duration was not proportional to the amplitude of the N1 ($N1_{MAXAMP}$) in the case of ACEP (($W_{N1}$ = -0.016 × $N1_{MAXAMP}$ + 38,27, $R^2$=0.128, n=8 patients/values) and was proportional for DCR but still with a low proportion of the variance explained by the data ($W_{N1}$ = 0.041 × $N1_{MAXAMP}$ + 38,46, $R^2$=0.455, n=9 patients/values) suggesting distinct shapes. When all the data are considered, $R^2$ remain very low: 0.0934, 0.152 and 0.174 for the ACEP (n=29), the DCR (n=55) and all data together, excluding the robustness of any proportionality between $W_{N1}$ and $N1_{MAXAMP}$.

### 3.4. After positivity in ACEP vs. monotonic decrease of the N1 response for DCR

The area under the ACEP curves between 40 and 100ms is positive and greater than that measured for the DCR. The DCR generally still remain negative after 40ms. Furthermore, this after-positivity in ACEP is not monotonic and presents a positive slope followed by a negative slope with a peak between 50 and 80ms. Conversely, for the same time interval, the slope of the DCRs remain monotonous and positive.

### 3.5. Time-Frequency content analysis

Gamma activity around 50Hz and averaged for five time-windows (20 to 40ms) was superior for DCR than for ACEP (paired t-test: t=3.50, p= 0.0081; see Figure 3). We could not observe any other statistical differences, despite visual differences in the time-frequencies color maps.

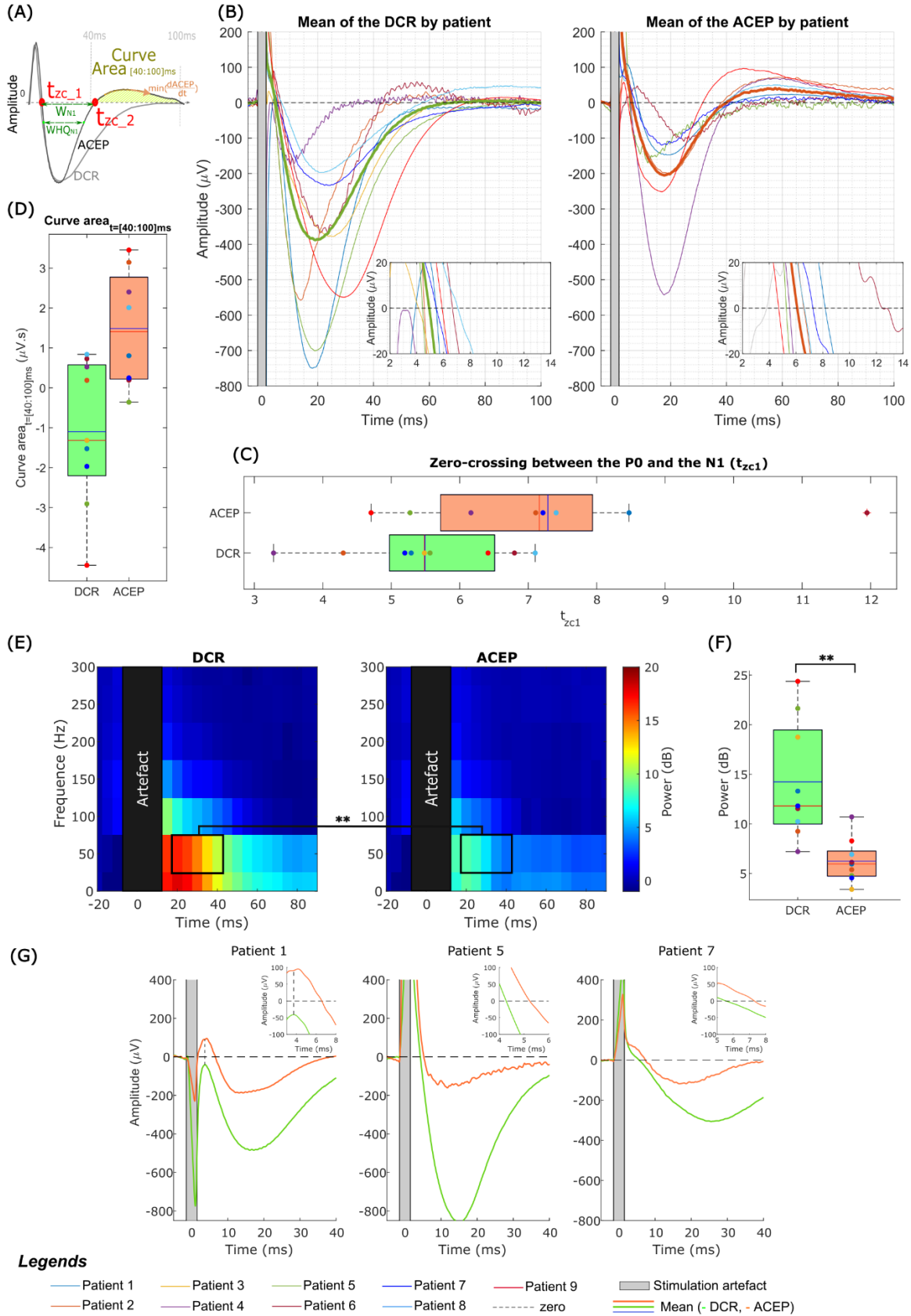

**Figure 3.** Metrics (A) and Results (B-F). (B) Averaged traces of DCR and ACEP for each patient. Green and red thick traces represent the mean for all patients of DCR and ACEP respectively. Inset shows the early time window (2.5-15 ms) zooming on the phase delay observed for the first zero crossing of the traces (beginning of the N1 component); (C) Boxplots of the first zero crossing (t_zc1, i.e. beginning of N1 component), (D) the signed areas under the curves between 40 and 100 ms and (F) the power centered on the 50Hz activity for the 20 to 40 ms windows pooled together, for DCR (green) and ACEP (red), including mean, median and individual values (see legend at the bottom). (E) Averaged frequency spectrum for DCR and ACEP. (G) ACEP and DCR sharing the same recording sites. Patient 1: 2.5 mA, the observed delay between P0 peaks is not greater for DCR than for ACEP; Patient 5: 2 mA, DCR tzc1 = 4.25 $\pm$ 0.3 ms, estimated distance between the stimulation and recording site 2.1 cm ACEP tzc1 = 4.96 $\pm$ 0.38 ms (p =6.86e-5, 2229); Patient 7: 1.5 mA, DCR tzc1 = 5.40 $\pm$ 1.91 ms, ACEP estimated distance between the stimulation and recording site 2.3cm  tzc1 = 7.28 $\pm$ 1.64 ms (p =1.71e-7, 26422).

**Discussion**

The objective of this study was to examine the distinctions between DCR and ACEP. Specifically, we hypothesized that certain electrophysiological events would manifest with delays for ACEP, and that intra-cortical activity would also exhibit variations between DCR and ACEP. These distinctions are attributed to the application of DES at different sites (namely, at the level of the cortical surface and within the same gyrus for DCR, and on the white matter tracts in the case of ACEP).

Despite the markedly distinct stimulation sites, these two typical types of responses exhibit common features. This strongly indicates that identical electrophysiological activities are elicited by direct electrical stimulation (DES), whether it is applied within the cavity on the white matter tracts or on the cortical surface and gray matter. Although unexpected, this observation aligns with the mechanisms of action of DES, which initially activates large neuronal elements, particularly the largest axons (given their lower excitability threshold,(Blair and Erlanger, 1933; Nowak and Bullier, 1998a, b).

This observation is further substantiated by the similarity in shapes between cortico-cortical evoked potentials (CCEP) and the two types of responses investigated in this study (NB: graphically, temporal scales and polarity may differ in these CCEP studies (Rossel et al., 2023), but waveforms remain strongly similar for typical responses). The typical waveform comprises an early response, P0 (activation of afferent and efferent axons within the cortical column), followed by a later response, N1 (summation of excitatory postsynaptic potentials, EPSP), potentially succeeded by some subsequent responses that are more integrated (Rossel et al., 2023). This classical waveform appears to be consistently observed irrespective of the DES site.

As anticipated, specific differences between DCR and ACEP are observable. In the case of ACEP, the application of DES to the subcortical white matter tracts, primarily the gyral ones in this study, introduces a delay in the initiation of the N1 component, measurable even for a few milliseconds.

It is important to emphasize that these differences in delays are difficult to highlight because DES can be applied or overflow onto an adjacent gyrus in the case of DCR. In this case, it is not a true DCR which is generated but a short-range CCEP which can propagate through the U fibers and can superimpose to the actual DCR recorded on the electrode. These CCEP will have delays and similar forms to the ACEP. It is quite difficult to have strict control over this possible bias which plays against our hypotheses.

Despite imprecisions in estimating the actual distances covered by AP in ACEP, the observed delays, providing estimates of the conduction velocities calculated in this study for each patient, align with those reported in the literature. For instance, for long range ACEP, Yamao et al. (2014) reported a total conduction latency of 12.8 ms for the whole arcuate fasciculus, whose length can be estimated to about 10 cm. Our estimate (5.9 m.s$^{-1}$) is in accordance with this result (7.8 m.s$^{-1}$) and remain in the range of theoretical values. It is well established in primates, including humans, that there exist large white matter fibers (up to 12 μm), such as the corticospinal tract (Lassek 1954, Firmin et al. 2014), which are highly myelinated and likely contribute significantly to the D-wave. However, resections involving the pyramidal tract (and optic tracts) are typically not performed, but due to these resections, radial fibers projecting to certain subcortical structures are cut. This is why we referred to the data of Liewald et al. (2014), which focuses on cortico-cortical associative pathways with much smaller diameters. Nevertheless, similar to the D-wave (Kraskov et al., 2020; Lemon, 2008), ACEPs, owing to the properties of DES, may reflect the activation of the fastest and largest fibers, akin to how the D-wave predominantly represents the fastest 2% of fibers in the corticospinal tract (Lemon 2008, Lassek 1954). Given this, the onset latency of N1 may need to be predicted based on the fastest fibers rather than an average.

On the contrary, the application of DES to the cortical surface in the case of DCR appears to elicit more pronounced activation of smaller and slower intra-cortical axons within the cortical layer surrounding the pyramidal neurons (refer to Figure 1). This tangential intra-cortical activation could bring about alterations in the later components of the N1 for DCR. It might induce the summation of small radial dipoles (e.g., excitatory postsynaptic potentials, EPSP) with diverse temporal and spatial distributions around the pyramidal cells, resulting in a slower relaxation phase of the N1. Notably, consistent with this interpretation, the impact of DES, attributed to the diminutive size of non-myelinated intra-cortical axons, may be confined to the superficial layers of the cortex.

One may also suggest that differences in the relaxation shape of the N1 between ACEP and DCR could be due to an increased spatial effect of DES for DCR. Hence, wider and greater N1 should be observed for DCR. However, this observation and this interpretation was not verified when correlating the duration of the N1 with its maximum amplitude.

We were also unable to demonstrate a greater and sustained gamma activity for DCR in the N1 phase associated with this the whole relaxation period as an illustration of the desynchronized summation of some radial dipoles intra-cortically. Significant activity in the 50Hz band was observed only at specific time windows, from 20 to 40 ms (with a step of 5 ms), notably in the proximity of the N1 maximum amplitude, for DCR. While intriguing, this result might be influenced by the increased N1 activity observed for DCR. Conversely, in the case of ACEP, the after-positivity could be related to inhibitory processes (Usami et al., 2015) that spatially restrict the spread of intra-cortical activation around the cortical columns.

Based on a 'convergent' paradigm in which DES is performed from different sites and EPs are recorded at the same site, Miller et al (2023) argued that "there is no single canonical CCEP response shape or feature, even when measuring from a single electrode." Their findings suggest that different waveforms may be specifically mapped to distinct cortical and subcortical anatomies. They also propose that the P0 component could be artifactual.

However, since the pioneering work on " the spread of activity in the cerebral cortex" by Adrian (1936) followed by numerous studies (Bishop and Clare, 1953; Chang, 1951; Clare and Bishop, 1955; Eccles, 1951; Goldring et al., 1961a; Li and Chou, 1962; Purpura et al., 1957; Sugaya et al., 1964; Tasaki et al., 1954) and more recently the work of Professor Matsumoto's team (Matsumoto et al., 2017, 2012, 2007, 2004; Usami et al., 2015; Yamao et al., 2014b), we have very frequently observed a canonical form in the initial part of the DCR, ACEP, and CCEP with a P0 (when it can be measured, Goldring et al., 1994; Goldring et al., 1961a, 1961b; Landau, 1956; Landau and Clare, 1956; Sugaya et al., 1964), an N1 which corresponds to Adrian's negative surface potential, and possibly, but not always, an after positivity and an N2. To identify these similarities, it is sometimes necessary to pay attention to the polarity of the representation (e.g., negative up vs. positive up) but above all to rescale within the correct time window (the time window representations of EPs often depend on the DES frequency used).

This frequently observed form, referred to as "canonical," persists despite variations in parameters and DES sites for constant recording sites (as particularly practiced by Matsumoto's group in epilepsy research). Notably, without explicitly naming it, Matsumoto's group has data that mimic Miller et al.'s convergence paradigm by stimulating multiple locations and recording with a mesh of large ECoG electrodes in epileptic patients. They clearly observe canonical forms.

Specifically, Yamao et al. (2014) also observed ACEPs with similar shapes to CCEPs but with different latencies when stimulating the arcuate fasciculus (see Yamao et al., 2014).

Variations may occur, and we have also observed some (see Rossel et al., 2023, Figure 4.). We hypothesized (Rossel et al., 2023; see discussion) that these variations could be due to the combination of distinct macroscopic dipoles or generators with different orientations.

Miller et al. also seem to show minimal interest, if any, in the initial segment of the signal, particularly in P0, which remains a highly debated component. This lack of attention primarily stems from methodological reasons.

In terms of recording, we collect our data at 19.2 kHz, while most other teams record at around 2 kHz. Additionally, we record without an implemented hardware filter, unlike many other teams whose recording systems have such limitations. If hardware filtering is present, it is sometimes necessary to characterize the filter's response to electrical stimulation in vitro (e.g., in a saline solution) to potentially subtract it (Vincent et al., 2020). Finally, it is important to note that the alternation of DES polarity is crucial to confirm that the signal is physiological and does not change polarity with the stimulation (Vincent et al., 2017).

Consequently, these two specifications (very high sampling frequency and absence of a hardware filter), combined with the alternation of stimulation polarity, enable us to record the DES artifact very precisely and without distortion. This allows for easy removal of the artifact by interpolating between its start and end. Additionally, the high sampling frequency enables us to accurately identify the 50 Hz component and remove it by subtraction (see methods and supplementary material/appendix).

From the viewpoint of stimulation, which is often overlooked or neglected, we employ different methods compared to Miller et al. We stimulate at low intensities (1-3 mA) using a bipolar electrode with a contact diameter of 0.5 mm, whereas Miller et al. use significantly larger ECoG contacts with a diameter of 5 mm or sEEG contacts for much higher intensities (>6 mA). Consequently, it is possible that these larger DES surfaces and higher intensities increase the probability of inducing distinct macroscopic dipoles, compared to our more selective DES modalities.

In this context, there are also notable distinctions. During tumor surgery, resection involves cutting "radial" projection fibers within specific deep structures, potentially lowering the likelihood of generating dipoles at this level and limiting the combination of "canonical" cortical dipoles with different orientations or other dipoles. Conversely, in epilepsy cases, all fascicles remain intact, allowing for the exploration of connectivity over greater distances. Furthermore, craniotomy during tumor surgery spatially restricts the measurement of CCEPs with ECoG electrodes. Despite these differences, the DES methodologies employed by Miller et al. closely align with those used by Matsumoto et al. Figure 4 below summarizes these considerations.

Finally, Miller et al. also focus more on late events. However, as we move further away in time from the stimulation artifact, the initially evoked activity, which is highly synchronous, tends to "renormalize" towards a desynchronized spontaneous activity. Consequently, in this scenario, the network's response may be more susceptible to variations in state (e.g., pharmacological, excitability), the origin of DES, and other factors, rendering it more unstable (see Figure 4A below).

Regarding the interpretation of the P0 component, we contend, supported by several studies, that its origin is associated with the synchronous activation of white matter fascicles. Conventionally, brain electrophysiological activities recorded on the surface (e.g., EEG, ECoG) are understood to result from the spatio-temporal summation of Post-Synaptic Potentials (PSPs). While this holds true for spontaneous activities, it is highly debatable for evoked activities involving extensive synchronous activation of white matter tracts.

Sugaya et al. (1964) noted that "the several brief positive spikes which initiate DCR when the evoking stimulus is strong were attributed to the all-or-none discharge of neurone somata situated in the cortical depth;". Additionally, it is well-known that electrical stimulation initially activates larger neural axons rather than smaller ones or other neural elements such as the soma (Blair and Erlanger, 1933; McNeal, 1976; L G Nowak and Bullier, 1998a, 1998b; Rattay, 1999; Ruch TC and Patton HD, 1982).

Moreover, geometrically, it is implausible to synchronize and summate PSPs whose peak occur within 10 ms to generate an overall P0 component with a peak occurring before 10 ms. However, high synchronization and summation of spikes or action potentials could feasibly result in a P0 peak around 5-8 ms.

Thus, P0 bears resemblance to an electroneurogram (ENG) of the brain, akin to measurements made on peripheral nerves (with comparable amplitudes in the range of a few tens of µV). Its measurement is delicate because, corresponding to the summation of very brief action potentials, variations in the orientation of white matter fibers can cancel out this summation. It is also important to refrain from using "P1" terminology, even if remotely measured, as this term should instead characterize a polarity inversion of N1 associated with a dipole polarity change, such as in a sulcus, for instance (see Rossel et al., 2023).

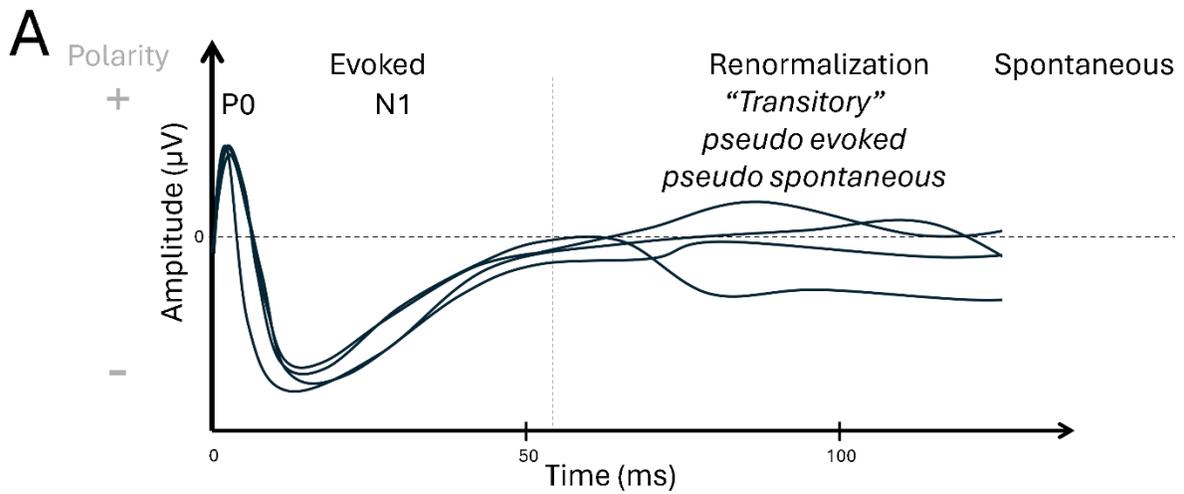

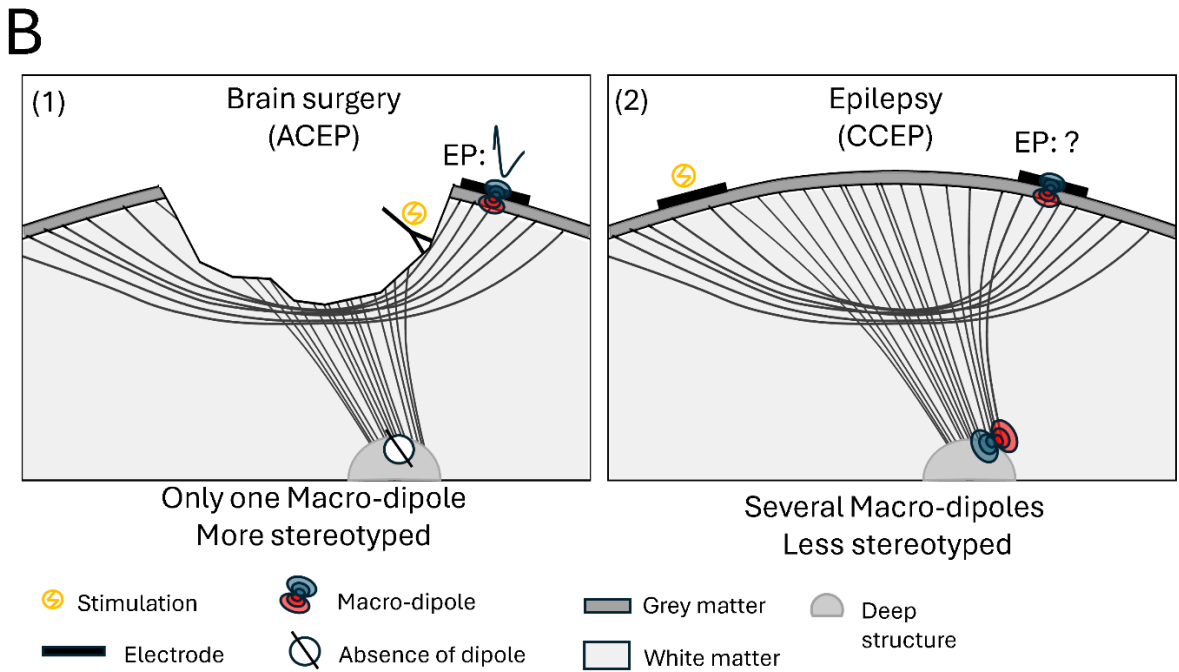

**Figure 4.** (A) EP Waveforms illustrating the canonical form observed in recorded signals, with N1 and P0 as principal components elicited by stimulation followed by signal renormalization upon return to spontaneous activity. (B). Explanation of waveform variations due to the combination of distinct macroscopic dipoles or generators with different orientations. (1) Brain surgery case: the stimulation applied in the cavity (shown in yellow) generates dipoles in the cortical region, resulting in a more stereotyped EP. (2) Epilepsy case: larger DES applied to the cortex generates distant dipoles, which may increase the likelihood of inducing distinct macroscopic dipoles, resulting in a less stereotyped EP.

One of the limitations of this study is the relatively short distances involved due to the reduced craniotomy. In many cases, ACEPs are triggered in the gyral white matter, indicating they originate from shallow sites. Additionally, it is challenging to control the spatial resolution or selectivity of the DES, particularly due to the orientation of the probe (Schlosser-Perrin et al., 2023), and because the post-operative positioning of the stimulation probe is also imprecise. Consequently, certain ACEPs at very short distances could also result in direct cortical stimulation from below. Although we have attempted to control for these various factors, in clinical practice, some confounding variables may persist. Therefore, it is important to study ACEPs over longer propagation distances.

In conclusion, a canonical form appears in DCR, ACEP, and CCEP because the recorded response primarily depends on the cortical output level at the recording site. Integrative stages associated with input stimulation play a minimal role as their effects are desynchronized, resulting in weak remote summation. However, it seems feasible to distinguish between ACEP and DCR based on a delay at the onset of the N1 component and the divergence in the shape of the late components (>40 ms after the DES artifact). These findings carry significant implications for enhancing our comprehension of the impacts of DES and refining its coarse grain modeling for brain mapping. It seems important to carefully assess subtle changes in the waveform of evoked potentials, (even though some canonical waveforms exist), to better understand the effects of DES in brain and evoked electrogenesis.

**Declaration of competing interest**

The authors declare that they have no known competing financial interests or personal relationships that could have appeared to influence the work reported in this paper.

**Acknowledgments**

Research supported by French ''Association pour la Recherche sur le Cancer" (ARC-France, project ARCPGA12019110000940_1572), the LabEx NUMEV project (ANR-10-LABX-0020) within the I-Site MUSE (ANR-16-IDEX- 0006); and INRIA through an associate team between France (FB) and Japan (RM). RM was supported by a Kobe University Strategic International Collaborative Research Grant (Type B Fostering Joint Research) between Japan (RM) and France (HD, FB), and JSPS KAKENHI 22H02945, 22H04777. We appreciate the insightful comments from both reviewers.

**Authors contributions**

CT, OR, FSP, FB collected and processed the data. HD, SN, EM performed the Neurosurgeries and participated to the collection of data. CT, FB wrote the first version of the manuscript. All authors participated to discussions and corrections for the final preparation and edition of the manuscript. FB and HD supervised the research.